\documentclass{article}
\usepackage{authblk}
\usepackage{ae}
\usepackage[T1]{fontenc}
\usepackage[ansinew]{inputenc}
\usepackage{amsmath}
\usepackage{amssymb}
\usepackage{color}
\usepackage{graphicx}
\usepackage{longtable}
\usepackage{latexsym}
\usepackage{stmaryrd}
\usepackage{amsfonts}
\usepackage{amsbsy}
\usepackage{mathrsfs}
\usepackage{psfrag}
\usepackage{enumerate}
\usepackage{MnSymbol}
\usepackage{vmargin}
\setmarginsrb{3cm}{2cm}{3cm}{2cm}{1cm}{1cm}{1cm}{1cm}
\usepackage{hyperref}
\hypersetup{colorlinks=false}

\usepackage{amsmath,amssymb,calc,amsfonts}
\usepackage{latexsym}

\usepackage{graphicx,calc,epsfig}
\def\ut#1{\rlap{\lower1ex\hbox{$\sim$}}#1{}}
\def\sdpb#1{\rlap{\lower1.5ex\hbox{$\Leftarrow$}}{#1}}

\newcommand{\R}{\mathbb{R}}
\newcommand{\Z}{\mathbb{Z}}
\newcommand{\be}{\nopagebreak[3]\begin{equation}}
\newcommand{\ee}{\end{equation}}
\newcommand{\bee}{\nopagebreak[3]\begin{equation*}}
\newcommand{\eee}{\end{equation*}}
\newcommand{\ba}{\nopagebreak[3]\begin{eqnarray}}
\newcommand{\ea}{\end{eqnarray}}
\newcommand{\baa}{\nopagebreak[3]\begin{eqnarray*}}
\newcommand{\eaa}{\end{eqnarray*}}
\newcommand{\la}{\label}
\DeclareFontFamily{U}{rsfs}{}         
\DeclareFontShape{U}{rsfs}{m}{n}{<5> rsfs5 <6><7> rsfs7          %
  <8><9><10><10.95><12><14.4><17.28><20.74><24.88> rsfs10}{}     %
\DeclareMathAlphabet{\mathfs}{U}{rsfs}{m}{n}                     %
\newcommand{\mfs}[1]{\mathfs {#1}}                               %
\newcommand{\va}{\scriptscriptstyle}

\newcommand{\n}{{\nonumber}}

\newcommand{\sH}{{\mfs H}}

\newcommand{\Hk}{{\sH}_{kin}}

\newcommand{\g}{\mathfrak{g}}
\newcommand{\su}{\mathfrak{su}}

\newcommand{\tr}{\mathrm{tr}}

\def\i{i}

\begin{document}

\title{CFT/Gravity Correspondence on the Isolated Horizon}

\author[1]{Amit Ghosh\thanks{amit.ghosh@saha.ac.in}}
\author[2]{Daniele Pranzetti\thanks{daniele.pranzetti@gravity.fau.de}}

\affil[1]{{\it Saha Institute of Nuclear Physics}, 

1/AF Bidhan Nagar, 700064 Kolkata, India.}

\affil[2]{{\it Institute for Quantum Gravity}

University of Erlangen-N\"urnberg (FAU), 

Staudtstrasse 7 / B2, 91058 Erlangen, Germany.}

\sloppy
\maketitle

\pagestyle{plain}

\begin{abstract}
A quantum isolated horizon can be modeled by an $SU(2)$ Chern-Simons theory on a punctured 2-sphere. We show how a local 2-dimensional conformal symmetry arises at each puncture inducing an infinite set of new observables localised at the horizon which satisfy a Kac-Moody algebra. By means of the isolated horizon boundary conditions, we represent the gravitational fluxes degrees of freedom in terms of the zero modes of the Kac-Moody algebra defined on the boundary of a punctured disk. In this way, our construction encodes a precise notion of CFT/gravity correspondence. The higher modes in the algebra represent new nongeometric charges which can be represented in terms of free matter field degrees of freedom. When computing the CFT partition function of the system, these new states induce an extra degeneracy factor, representing the density of horizon states at a given energy level, which reproduces the Bekenstein's holographic bound for an imaginary Immirzi parameter. This allows us to recover the 
Bekenstein-Hawking entropy formula without the large quantum gravity corrections associated with the number of punctures.

\end{abstract}

\sloppy
\maketitle
\newpage{}

\section{Introduction}\la{sec:Intro}

The notion of entropy for a black hole (BH) associated to its area \cite{Bekenstein, Hawking} led to a holographic picture of the black hole horizon carrying some information with a density of approximately 1 bit per unit Planck sized area. At the same time, the holographic picture raised an issue of the microscopic origin of such a large number of degrees of freedom on the horizon. The expectation that on one hand, the nature of these bits of information accounting for the entropy depends on the quantum structure of space-time and on the other, the finiteness of the black hole entropy hints at the discreteness of space-time at the Planck scale are both realised in the loop quantum gravity (LQG) framework. Applying the LQG quantisation techniques to {\it isolate horizons} (IH---a local definition of a black hole \cite{ABCK, SU(2)} horizon) these degrees of freedom have been identified with the polymer-like excitations of the gravitational field encoded in the spin network states and spanning the kinematical 
Hilbert space of the theory, piercing the horizon at a collection of points called {\em punctures}. 

More precisely, the bulk theory gets coupled to a boundary $SU(2)$ Chern-Simons theory on the IH by inducing conical singularities in the curvature of the Chern-Simons gauge fields at the location of punctures where the edges of the spin network pierce the horizon. The dimension of the resulting Chern-Simons Hilbert space of the punctured two-sphere has been computed in different statistical mechanical ensembles and shown to generate a leading term linear in the IH area \cite{Lewa, GM, Counting2, Counting, Livine-Terno} (see \cite{Review} for a review) which matches the Bekenstein-Hawking formula for a specific numerical real value $\gamma_0$ of the Barbero-Immirzi parameter $\gamma$. However, this feature of the standard LQG black hole entropy calculation has raised   concern in the past years.  The reason being that it is not clear at all what the role of $\gamma$  in the semi-classical limit of the theory should be (if any) and, hence, fixing its value via a match with the numerical factor in the Bekenstein-Hawking formula (the famous $1/4$ is a semiclassical result of a QFT on curved space-time calculation) is not well justified at this stage.

 More recent calculations have shown that an exact matching up to the numerical factor in the semi-classical result is obtained straightforwardly for any value of $\gamma$ once  the Unruh temperature is introduced in the IH partition function as an external (semi-classical) physical input  \cite{AP, Radiation}. However, a quantum gravity correction in the entropy formula appears in the form of a chemical potential term associated with an additional quantum hair represented by the number of punctures. Later on, a proposal for the removal of 
this extra term via the 
inclusion of a new degeneracy factor in the partition function associated with matter degrees of freedom has been proposed in \cite{GNP}. This raises the important question of the role of matter in the LQG calculation of black hole entropy which has so far been largely overlooked. 

Together with the introduction of the Unruh temperature for a local stationary observer near the horizon in the canonical (and grand canonical) ensemble treatment, another recent observation in the micro-canonical ensemble framework has provided a new perspective in order to get rid of the unphysical numerical constraint on the Barbero-Immirzi parameter. In \cite{Complex} it was observed that, by doing an analytic continuation of the Verlinde formula for the IH Hilbert space dimension to imaginary $\gamma$, the Bekenstein-Hawking entropy is recovered in the large spin limit. While this might seem like just a different numerical constraint on the Barbero-Immirzi parameter, $\gamma=\pm i$ plays a special role in the theory. In fact, this choice of $\gamma$ corresponds to the original Ashtekar self-dual variables \cite{Ash-con} and these are physically preferred with respect to the real ones since only the complex Ashtekar connection has the right transformation properties under space-time diffeomorphisms \cite{
Samuel} and can be derived from a manifestly covariant action \cite{JS}. However, constructing the kinematical Hilbert space of the theory with the complex variables turned out to be very complicated and they were originally dismissed in favor of the real $SU(2)$ variables, which allowed to generalize in a quite straightforward way techniques previously developed in the context of Yang-Mills theory.  

Therefore, the new perspective that has emerged is to regard real $\gamma$ as a regulator for the theory to perform a well defined (BH entropy) calculation and then perform an analytic continuation to imaginary Barbero-Immirzi parameter at the end in order to recover physical results. In fact, the physical relevance of $\gamma=\pm i$ in the framework of isolated horizons was pointed out in  \cite{Temp}, where a geometric notion of IH temperature was derived form the microscopic theory without any external semi-classical input and the thermality of the horizon was shown to be intimately related to the passage to the Ashtekar self-dual variables. In this way, a connection between the observations of \cite{AP} and \cite{Complex} was established. Moreover, while the result of \cite{Donnelly} provides an entanglement entropy interpretation to the counting performed in \cite{ABCK, Lewa}, in \cite{Temp} it was also shown that the statistical entropy obtained in \cite{AP} in the canonical ensemble matches the 
entanglement entropy of the thermal density matrix associated to the quantum IH, once $\gamma$ is taken purely imaginary (see also \cite{Perez} for the connection between statistical and entanglement entropy). This tension is probably related to the different nature of degrees of freedom traced over in the two cases and, hence, of the resulting density matrices. Further evidence for the special role of $\gamma=i$ in the semiclassical limit was pointed out also in \cite{BodeN}, through the consistency of the spin foam amplitude large spin limit with the (discrete) Einstein-Hilbert action of general relativity.

Therefore, in the past few years, an alternative perspective has emerged with respect to the original treatment of \cite{ABCK}, which might provide important physical insights in the full theory too. However, the new picture is not well understood and fully self-consistent yet. In particular, it would be desirable to find the origin of the ad-hoc extra degeneracy factor introduced in the IH canonical partition function in \cite{GNP} within the LQG structures (notice that such factor is crucial in order to justify the large spin limit taken in \cite{Complex}), without relying on semi-classical arguments. This would make the new point of view on BH entropy calculation in LQG more self-consistent and might open a new way to couple matter d.o.f. to the theory, an issue which is far from being successfully addressed and understood. 

Moreover, the connection between the analytic continuation of $\gamma$ performed in \cite{Complex} and the presence of the extra degeneracy factor introduced in  \cite{GNP} is not clear yet. Both approaches seem to give the correct answer, however the analysis in \cite{Complex} is performed in the micro canonical ensemble by keeping the set of external puncture fixed, while \cite{GNP} works in the grand canonical ensemble and a sum over spins is performed. In the latter case, two fundamental ingredients where introduced by hand: the Unruh temperature for a local stationary observer near the horizon and the new degeneracy factor. As pointed out above, the role of imaginary $\gamma$ for an actual derivation of the IH (geometric) temperature was first elucidated in \cite{Temp}; here we show that $\gamma= i$ also plays a fundamental role for the derivation of such a new degeneracy factor, matching the Bekenstein bound. However, the role, if any, of this new degrees of freedom in the context of  \cite{Complex} is 
not clear yet and it deserves further investigation.

At the same time, given the current incompleteness of all fundamental approaches, alternative methods to compute black hole entropy have been developed which do not depend on the details of any quantum theory of gravity but just on symmetry arguments. This line of approach was pioneered by \cite{Carlip94} and improved by \cite{Strominger} who used the analysis of \cite{Brown} and applied the results of conformal field theory (CFT) to extremal black holes having ${\rm AdS}_3$ near-horizon geometry. In particular, by applying Cardy's formula \cite{Cardy} for computing the asymptotic density of states in the CFT, remarkable agreement with the Bekenstein-Hawking formula was found. Extension of this method to dimensions higher than three and black holes which are non-extremal was obtained right after by looking at a Virasoro sub-algebra of the surface deformation algebra at the horizon (i.e. for an effective 2-dim conformal field theory on the boundary) \cite{Carlip, Solodukhin, DAA}. The generality and strength 
of this method has motivated the conjecture that a two-dimensional dual conformal theory may emerge as an effective description of the near horizon degrees of freedom from any underlying quantum gravity approach to black hole entropy \cite{Carlip07}. However, another possibility could be that {\it local} conformal symmetry provides an exact description of the microscopic quantum gravity (and possibly also matter) degrees of freedom.

Thus, in the context of LQG, we confront two main questions: 1) Are there new degrees of freedom at the horizon, possibly related to matter, which were missed in previous counting and which can provide the correct degeneracy factor in the partition function in order to recover the Bekenstein-Hawking formula exactly? 2) Is there a dual 2-dim CFT symmetry lurking somewhere in the LQG description? 

The aim of this paper is to answer both questions. Explicitly, we show that the IH boundary conditions allow us to define a new set of horizon observables that satisfy the $SU(2)$ affine Kac-Moody algebra. We show that the generators of this affine algebra contains both the gravitational degrees of freedom, previously taken into account in the LQG counting, and some new ones which furnish a representation in terms of free matter fields. In this way, a dual local CFT description is explicitly realised at each puncture on the horizon. The conformal symmetry is then used to write a partition function of the quantum IH, taking into account also the new degrees of freedom. We show that a new degeneracy factor appears which exactly reproduces Bekenstein's holographic entropy bound for an imaginary Barbero-Immirzi parameter.  

Note that the well known map \cite{Witten} between a Chern-Simons theory on a punctured 2-sphere and the space of conformal blocks of an $SU(2)$ WZW theory has already been exploited to estimate the dimension of the IH Hilbert space (see \cite{KM} and \cite{SU(2)}). However, the full extent of implications of a local conformal symmetry on the horizon has never been investigated in the previous literature of quantum isolated horizons.   

The plan of the paper is as follows. In section \ref{sec:Kac-Moody} we introduce the new set of observables spanning a Kac-Moody algebra and define a unitary irreducible highest weight representation. In section \ref{sec:Virasoro}, by means of the Sugawara construction, we define the associated Virasoro algebra, specifying its central charge and the zero mode that plays the role of an energy operator. In section \ref{sec:vertex} we show how the vertex operators of the Verma module associated with the Virasoro algebra has an interpretation in terms of holonomies; this allows us to provide a LQG description of CFT states. We also provide a free field representation of the new degrees of freedom. In section \ref{sec:Entropy} we introduce the CFT partition function and use the modular invariance to show how the new degrees of freedom lead to the appearance an holographic entropy bound which is consistent with the semiclassical derivation. Section \ref{sec:Conclusions} contains conclusions and some comments on 
our results. Some complementary material is added in the two appendices. 


\section{CFT/Gravity}

As briefly recalled in the Introduction, the isolated horizon boundary conditions lead to the picture of a quantum horizon as described by a Chern-Simons theory on a punctured 2-sphere. By momentarily focusing on a single puncture, we now show how this implies the existence of a local conformal symmetry.

\subsection{Kac-Moody algebra}\la{sec:Kac-Moody}

The $SU(2)$ Chern-Simons (CS) action on a solid cylinder $D\times \R$ is
\be\la{Action-CS}
S_{\rm CS}=\frac{k}{2\pi}\int_{D\times \R} \tr{[A\wedge dA+\frac{2}{3}A\wedge A\wedge A]}\,,
\ee
where the connection $A=A^i_\mu dx^\mu \tau_i$ has curvature $F(A)=dA+A\wedge A$ and the generators $\{\tau_i\}$ form a basis
of the $\su(2)$ Lie algebra with normalization $\tr{(\tau_i\tau_j)}=-(1/2)\delta_{ij}$ and $[\tau_i,\tau_j]=\epsilon_{ijk}\tau_k$. The phase space of the theory \eqref{Action-CS} is described by the equal time Poisson brackets
\be\la{PB}
\{A_a^i(x), A_b^j(y)\}=\delta_{ij}\epsilon_{ab} \frac{4\pi}{k}\delta^2(x-y),~~~a,b=1,2;~x^0=y^0
\ee
with the convention $\epsilon_{012}\equiv\epsilon_{12}=-\epsilon_{21}=1$.
Interaction with a static non-abelian charged source at a point $p$ in $D$ with the CS field is encoded in the interaction action \cite{Witten, Elitzur}
\be\la{Action-int}
S_{\rm int}=\lambda_j\int_{c} \tr{[\tau_3(\Lambda^{-1}d \Lambda+\Lambda^{-1}A \Lambda)]}\,,
\ee
where $c$ is a Wilson-line (a timelike path), $\lambda_j$ is a coupling constant in which the label $j$ refers to the spin-$j$ unitary irreducible representation of $SU(2)$ and $\Lambda$ is a group element in the fundamental representation. Let us expand the Lie algebra element $\Lambda^{-1}\tau_i\Lambda=L_{ik}\tau_k$ where the matrix $L$ is such that $L^{-1}=L^T$ and $\text{det}\,L=1$. In the Hamiltonian framework, variation of the two terms \eqref{Action-CS}, \eqref{Action-int} 
leads to the constraint
\be\la{con1}
F_{12}^i(A(x))=-\frac{2\pi}{k}S^i\delta^2(x-p)\,,
\ee
where $S^i=\frac{1}{2}\lambda_jL_{i3}$ is the conjugate variable to $\Lambda$ (the phase space of the source is $T^*(SU(2))$) in the sense that it satisfies the Poisson brackets
\be\la{S-algebra}
\{S^i,\Lambda\}=\tau^i\Lambda\,.
\ee
 
\begin{figure}[ht]
\centering
\includegraphics[scale=0.4]{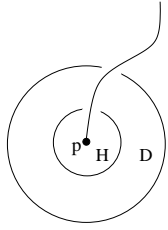}
\caption{Hole $H$ around the source $p$. $D$ is a disk around $H$ where the connection is flat. Its boundary $\partial D$ consists of two components, an inner boundary $\partial H$ and an outer boundary $B$.}
\label{hole}
\end{figure}

In presence of a point source the connection $A(x)$ is not well defined as $x$ approaches $p$ and, therefore, a regularization is needed \cite{Guadagnini}. This can be done by piercing a hole $H$ containing $p$ and eventually shrinking the hole to $p$ (see Figure \ref{hole}). Such regularization is the analog of the framing technique needed in the covariant context to solve the ambiguities appearing in the computation of the expectation values of the Wilson lines \cite{Witten}; it represents a standard procedure to deal with the conical singularity produced by a particle in 3-dim gravity \cite{Deser}.
Also in the context of isolated horizons, such regularization is not a novel feature of our approach. In fact, already in \cite{ABCK}, as an intermediary step to the construction of the horizon Hilbert space for a set of punctures on a 2-sphere, an analog set of paths was introduced. More precisely, in order to construct the quantum phase space consisting of generalized connections that are flat everywhere on the 2-sphere except at the punctures, it was necessary to introduce (at the classical level) some extra structure associated to certain equivalence class of paths. 
It was proven that such phase space is diffeomorphic to a $2(n-1)-$dimensional torus, with $n$ the number of punctures, and the Chern-Simons surface term in the symplectic structure of the classical theory can still be 
well-defined on this phase space, despite the distributional nature of the connections.

Moreover, the path $\partial H$ introduced by our regularization may not be smooth. The results will
hold for any closed path, which exists even on a quantum spacetime. In fact, one could also think
about these paths as spin network edges lying on the surface, as in the picture emerging, for instance, from the treatment of \cite{S}. Here the standard mixed quantization scheme for the bulk and boundary theories is replaced by a more uniform treatment of the horizon d.o.f. using LQG techniques and leading to an holonomy-flux algebra for the quantum IH. Along these lines, one could try to quantize the boundary theory as a 2+1 gravity theory coupled to point particles within the LQG framework\footnote{The equivalence between the quantization of 2+1 gravity in the Chern-Simons formulation and the BF formulation (using the LQG framework)  has been shown both in the $\Lambda=0$ \cite{NP2} and in the $\Lambda>0$ \cite{P} cases. It is then natural to expect that LQG techniques developed in the quantization of 3-dim gravity coupled to point particles be relevant to the quantization of isolated horizons entirely  within the  LQG  formalism. Work along these lines is in progress \cite{PS}.}. As shown in \cite{NP},  
in order to deal with the singular behavior of the connection at the location of the particle, a regularization scheme consisting of replacing the point particle by a small loop needs to be introduced. In the quantum theory then, one associates an $SU(2)$ spin representation and an intertwiner to each of these new boundaries, providing a generalization of spin network states. Thus, from a quantum geometry point of view, we find further evidence of the necessity to replace point particles with extended loops. 

Therefore, we see that the introduction of a new boundary around each puncture represents a standard and well motivated regularization procedure in order to quantize the IH boundary theory. The novel feature of our analysis  is to exploit the presence of these boundaries to its full extent, in order to make explicit and clarify the role of local conformal invariance on the quantum IH. In fact, in regularizing the point particles with small loops, the connection can now vary  along this finite boundary, giving rise to a quantum theory involving representations of the loop group. With only a puncture, i.e. a connection at a point, one obtains the ``old'' Hilbert space of conformal blocks. While with representations of the loop group an infinite tower of new dof now appears in the IH Hilbert space.
As explained in the Introduction, our main motivation for doing so is to show how the semiclassical Bekenstein-Hawking entropy formula can be recovered without large quantum corrections and for an imaginary Barbero-Immirzi parameter. At the same time, we will see how the new degrees of freedom, arising from such regularization, have an interpretation in terms of free matter fields, which may provide an alternative route to matter coupling in LQG and eventually a set-up to investigate Hawking radiation from a microscopic point of view.

Henceforth, after punching a hole $H$ around the puncture $p$, by means of non-Abelian Stokes' theorem the constraint \eqref{con1} can be rewritten as 
\be\la{con2}
\oint_{\partial H} A^i=-\frac{4\pi}{k} S^i\,,\quad\oint_{B} A^i=-\oint_{\partial H} A^i\,.
\ee
The Gauss' law takes the form 
\be\la{Gauss}
g(\alpha) = \frac{k}{2\pi}\int_D \tr{[\alpha F]}\approx 0\,,
\ee
where the test functions $\alpha^i$, $\alpha=\alpha^i\tau_i$, must vanish on $\partial H$ and $B$ in order for $g(\alpha)$ to be differentiable in $A$. The corresponding quantum operator $\hat g (\alpha)$ annihilates all physical states.

There exists a family of observables, localised on the boundary $\partial H$, that generate a Kac-Moody algebra \cite{Balachandran}. More precisely, given the test functions $\xi^{i}_N$ which vanish on $B$, the Kac-Moody generators are given by
\ba
q^{i}_{\va N}\equiv q(\xi^{i}_{\va N})\,,~~~ \xi^{i}_{\va N}|_{\partial H}=e^{\i N\theta}\tau_i\,,~~~\xi^{i}_{\va N}|_B=0\la{qH}\,,
\ea 
where $N$ is an integer, $\theta$ (mod $2\pi$) is an angular coordinate on $\partial H$ and 
\be\la{q}
q(\xi^{i})=\frac{k}{2\pi}\int_{D} \tr{[d\xi^{i}\wedge A- \xi^{i} A\wedge A]}\,.
\ee
The Poisson brackets of the $q(\xi^{i})$'s are
\be\la{algc}
\{q(\xi^{i}), q(\xi^{j})\}=q([\xi^{i},\xi^{j}])+\frac{k}{2\pi}\int_{\partial H} \tr{[\xi^{j} d \xi^{i}]}\,,
\ee
from which the commutation relations of the quantum operators $\hat q^{i}_{\va N}$ associated with these observables are given by
\be\la{KM}
[\hat q^{i}_{\va N}, \hat q^{j}_{\va M}]=\i\epsilon_{ijk}\hat q^{k}_{\va N+M}+\frac k2N\delta_{\va N+M,0}\delta_{ij}\,.
\ee
This shows that the observables $q^i_N$ generate a $SU(2)$ Kac-Moody algebra and are localised on the boundary $\partial H$. Now, after an integration by parts we can rewrite the observables \eqref{q} as
\ba\la{q2}
q(\xi)&=&\frac{k}{2\pi}\int_{D} \tr{[d\xi\wedge A- \xi A\wedge A]}
=\frac{k}{2\pi}\int_{\partial H} \tr{[\xi A]} - \frac{k}{2\pi}\int_{D} \tr{[\xi F]}\n\\
&=&\frac{k}{2\pi}\int_{\partial H} \tr{[\xi A]}\,,
\ea
where in the last step we have put $F_{12}(A)=0$ in $D$. It follows that
\be\la{qH2}
q^{i}_{\va N}=\frac{k}{2\pi}\oint_{\partial H} e^{\i N\theta}\tr{[\tau_i A]}=-\frac{k}{4\pi}\oint_{\partial H} e^{\i N\theta} A^i\,,\qquad q^i_0=S^i\,.
\ee

The choice of the test functions \eqref{qH} entering the definition of the new observables $q^i_{\va N}$ is motivated by the fact that the Virasoro generators constructed out of these Kac-Moody generators in a standard way (see Section \ref{sec:Virasoro}) correspond exactly to the generators of infinitesimal diffeomorphisms of a vector field tangent to the boundary loop---this is a reflection of the fact that the Virasoro algebra is the algebra of the diffeomorphisms on the circle. Then, the presence of a central extension in the algebra \eqref{KM} implies the appearance of  new states no longer annihilated by all of the Virasoro generators (see eq. \eqref{con3} below) associated to those diffeomorphisms: some degrees of freedom previously considered as gauge are now turned into physical ones due to the presence of a new boundary at each puncture. Therefore, the precise choice \eqref{qH} allows us to implement Carlip's proposal \cite{Carlip07} for the origin of BH entropy in an LQG context\footnote{Notice that already the original treatment of \cite{ABCK, 
Lewa} can be interpreted along these lines. In fact, the presence of point particles on the IH 2-sphere breaks those diffeomorphisms that exchange two or more of them.}. Namely, the LQG quantization of the bulk fluxes (identified with the r.h.s. of \eqref{con1} \cite{ABCK, SU(2)}) induces conical singularities in the strength of the Chern-Simons connection field at the location of the punctures;  in order to regularize this singular behavior, we blow the point particle to a finite loop, which we eventually srink back to zero; associated to these new boundaries there is a new set of  observables, corresponding to the diffeomorphisms along such loops, whose algebra now contains a central extension;  as a consequence, 
some of ``would-be pure gauge'' degrees of freedom can now contribute to BH entropy.

Let us now show how the Kac-Moody generators correspond to the modes of the holomorphic Chern-Simons connection field.
On a complex $z$-plane, an arbitrary holomorphic primary field $\phi(z)$ of weight $h$ admits a mode expansion (here, $z=\epsilon\,e^{i\theta}$ where $\epsilon>0$ is the radius of $H$) 
\be\la{primary}
\phi(z)=\sum_{N\in\,\Z}\phi_{\va N} z^{-N-h}
\ee
where the modes $\phi_N$ are given by
\be
\phi_{\va N}= \oint \frac{dz}{2\pi i}\,z^{N+h-1}\phi(z)\,.
\ee
The regularity of $\phi(z)|0\rangle$ at $z=0$ where $|0\rangle$ is the $SL(2,\R)$ invariant vacuum requires $\phi_{\va N}|0\rangle=0$ for $N\geq-h+1$. We can write the modes (\ref{qH2}) as an integral over the $z$-plane as
\be
 q^{i}_{\va N}=-k\oint \frac{dz}{2\pi i}\,z^N A^i(z)\,.\label{qin}
\ee
The above formula actually defines the holomorphic connection $A^i(z)$ as a primary field of weight 1, which admits the mode expansion
\be
A^i(z)=\frac{1}{k}\sum_{N\in\,\Z} q^{i}_{\va N}z^{-N-1}\,.\label{aiz}
\ee
An analogous expansion can be obtained for the anti-holomorphic part of the connection in terms of modes on the boundary $\partial H$. Unitary representations of the Kac-Moody algebra require the hermiticity condition
\be
\hat q^{i\dagger}_{\va N}= \hat q^{i}_{-\va N}\,.
\ee
%

The Kac-Moody generators \eqref{qH2} acts on the connection $A$ through canonical transformations. Since $\{q^i_N,A\}=d\xi^i_N+\epsilon_{ijk}\xi^j_NA^k$, the change does not modify the flux through the surface enclosed by $\partial H$ (the Poisson bracket of $q(\xi)$ with $\oint_{\partial H}A$ vanishes). Given a solution to \eqref{con2}, like for instance a blip $A_\theta=-\frac{2\pi }{k}S^i\delta(\theta-\theta_0)$ localised at $\theta_0$ on $\partial H$, one can generate other solutions having the same flux through these canonical transformations.

In order to find the irreducible representations of the $SU(2)$ Kac-Moody algebra \eqref{KM}, it is convenient to introduce a basis for the associated Lie algebra $\su(2)$ given by the generators of the Cartan sub-algebra $H^i$, $1\leq i\leq r$ where $r$ is the rank of the Lie algebra and the step operators $E^\alpha$ associated with the positive roots $\alpha$. In the case of (the complexification of) $\su(2)$, $r=1$, $\alpha=1$ and we can set 
\be
H^i\equiv H^3=\tau_3,~~~E^\alpha\equiv E^+=\tau_+=\tau_1+\i\tau_2\,,~~~E^{-\alpha}\equiv E^-=\tau_-=\tau_1-\i \tau_2\,.
\ee
In this basis the generators of the $SU(2)$ Kac-Moody algebra are
\be
\hat H^3_{N}=\hat q(e^{\i N\theta}\tau_3)\,,\qquad\hat E^{+}_{N}=\hat q(e^{\i N\theta}\tau_+)\,,\qquad
\hat E^{-}_{N}=\hat q(e^{\i N\theta}\tau_-)\,.
\ee
The affine Kac-Moody algebra \eqref{KM} can then be rewritten as
\ba
&&[\hat H^3_{N}, \hat H^3_{M}]=\frac k2N\delta_{N+M,0}\,,\n\\
&&[\hat H^3_{N}, \hat E^{\pm}_{M}]=\pm\hat E^{\pm}_{\va N+M}\,,\n\\
&&[\hat E^{+}_{N}, \hat E^{-}_{M}]=2\hat H^3_{\va N+M}+kN\delta_{N+M,0}\,,~~~~
[\hat E^{\pm}_{N}, \hat E^{\pm}_{M}]=0\,.
\ea
The hermiticity conditions now read
\be
\hat H^{3\dagger}_{N}=\hat H^3_{-N}\,,~~~\hat E^{+\dagger}_{N}=\hat E^{-}_{-N}\,.
\ee
Given a highest weight state of a unitary irreducible representation of $\su(2)$, namely a state $|v_j\rangle$ for which
\begin{align}\la{vj}
\hat H^3_{0}|v_j\rangle&=m_j|v_j\rangle\n\\
\hat H^3_{N}|v_j\rangle&=0\qquad\text{for}\;N>0\n\\
\hat E^{+}_{N}|v_j\rangle&=0\qquad\text{for}\;N\geq0\,,
\end{align}
all other states of a unitary irreducible highest weight module of the associated Kac-Moody algebra can be constructed by repeated application of the negative root operators $\hat E^{-}_{N}$ on $|v_j\rangle$. The states $|v_j\rangle$ form a representation of the finite dimensional algebra $\g=\{\hat q^{i}_{0}\}$, which is isomorphic to $\su(2)$. If the representation of $\g$ is irreducible then the representation of the associated Kac-Moody algebra is also irreducible.
Thus, the irreducible representations of the Kac-Moody algebra are characterised by the irreducible representations of $\su(2)$ and $k$.

\subsection{Virasoro algebra}\la{sec:Virasoro}

From the generators of a Kac-Moody algebra one can obtain the generators of a Virasoro algebra through the Sugawara construction \cite{Goddard, CFT}. The holomorphic stress-energy tensor $T(z)$ admits a Laurent expansion 
in terms of the Virasoro generators 
\be\la{T-exp}
T(z)=\sum_{N\in\,\Z} L_Nz^{-N-2}\,.
\ee
Clearly, $T(z)$ has the conformal dimension $h=2$ and the Virasoro generators are given in terms of the Kac-Moody ones by 
\be\la{L}
\hat L_{\va N} = \frac{1}{k+g}\sum_i\sum_{M\in\,\Z}\! :\hat q^i_{\va M} \hat q^i_{\va N-M}:
\ee
where $g$ is the dual Coxeter number, which is equal to $N$ for $SU(N)$ and $:\;:$ stands for the normal ordering, defined by
\be
:\hat q^i_{\va M}\hat q^i_{\va N}\!:\,=\begin{cases}
        \hat q^i_{\va M}\hat q^i_{\va N}\quad\text{if}\;N>M\\
        \hat q^i_{\va N}\hat q^i_{\va M}\quad\text{if}\;M>N\,.\end{cases}  
\ee
The normal ordering ensures finite matrix elements in a highest weight representation\footnote{The normal ordering prescription is relevant only for the zero mode $L_0$ since for the nonzero modes the Kac-Moody generators appearing in the expression \eqref{L} commute. As usual, the normal ordering is necessary to have finite energy values in a highest weight representation.}. The generators \eqref{L} generate diffeomorphisms of the boundary $\partial H$ and they satisfy the Virasoro algebra
\be\la{Vir}
[\hat L_{\va N}, \hat L_{\va M}]=(N-M) \hat L_{\va N+M} + \frac{c}{12}N(N^2-1)\delta_{N+M,0}\,,~~~~N,M\in\Z\,,
\ee
where $c$ is a central element, $[c,\hat L_{\va N}]=0$ for all $N\in\Z$, also called the {\em central charge}, which for a given Lie algebra $\g$, is given by
\be\la{c}
c=\frac{k~ {\rm dim}\,\g}{k+g}\,.
\ee
Among the generators \eqref{L}, the zero mode $\hat L_0$ plays a special role
\be\la{L0}
\hat L_0=\frac{1}{k+2}(\hat q_0^i \hat q_0^i+2\sum_{M>0}\hat q^i_{\va -M}\hat q^i_{\va M})\,,
\ee
which is also called the {\em energy operator}. We will see below that such a terminology is appropriate in our context also. In any highest weight representation the spectrum of $\hat L_0$ is bounded from below and there is a unique highest weight state $|v_j\rangle$ which satisfies
\be\la{con3}
\hat L_0 |v_j\rangle = \ell_j|v_j\rangle\,,\quad\hat L_{\va N} |v_j\rangle =0\,,~~\text{for all}\;N>0\,.
\ee
The eigenvalue $\ell_j$ of the zero mode is called the {\it conformal dimension} of the state $|v_j\rangle$. All other states in the highest weight representation (often called a Verma module) can be constructed by repeated application of $\hat L_{-N}$, $N>0$, on the highest weight state $|v_j\rangle$. Each irreducible unitary highest weight representation of the Virasoro algebra \eqref{Vir} is characterised by the pair $(c,\ell_j)$ where $c,\ell_j>0$. However, not all positive values of $(c,\ell_j)$ give rise to unitary representations. In our case, the relevant restrictions imposed by the unitarity are $c\geq 1$ and $\ell_j\geq 0$ (for $c<1$ only a discrete set of values for the central charge $c$ and the conformal dimension $\ell_j$ are allowed \cite{Goddard}).

\subsection{Holonomies as Vertex Operators}\la{sec:vertex}

\begin{figure}[ht]
\centering
\includegraphics[scale=0.4]{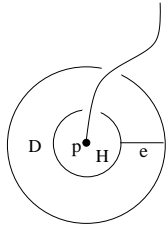}
\caption{An highest weight state can be constructed from the holonomy defined on the line $e$.}
\label{Disc2}
\end{figure}

The highest weight state $|v_j\rangle$ from which the Verma module associated with \eqref{Vir} is constructed, can be obtained from a vacuum state $|0\rangle$ by application of a vertex operator $\hat V_j$,
\be\la{vacuum}
|v_j\rangle=\hat V_j|0\rangle\,.
\ee
In the following, we are going to demonstrate how to construct the smallest representation. Define $Q=\hat q(-\theta\tau_3)$, where $\theta$ is the polar angle in $D$. The operator $Q$ is well defined in presence of a Wilson line \cite{Balachandran}. More precisely, given a Wilson line $e$ (see Figure \ref{Disc2}) from the inner boundary $\partial H$ to the outer boundary $B$ at a fixed angle $\theta=2\pi$, the polar coordinate $\theta$ is single valued in $D-e$ and (setting $k=1$ in \eqref{q})
\be\la{vertex}
Q=\lim_{H\rightarrow p}\frac{1}{2\pi}\int_{D-e} \tr{[-\tau_3d\theta\wedge A+\tau_3\theta A\wedge A]}=\int_e \tr {[\tau_3A]}\,,
 \ee
where we have used Gauss' law $\int_{D-e}\tr[\theta\tau_3F]\approx 0$ in the last step. 

In \eqref{algc} setting $k=1$ 
, we get
\be
[Q,\hat q(\xi^{i})]=-i\hat q([\theta\tau_3,\xi^{i}])-\frac{i}{2\pi}\oint_{\partial H} \tr{[\tau_3\xi^{i}]}d\theta\,,
\ee
from which, it follows that
\ba
&&[Q,\hat H^{3}_{N}]=\frac i2\delta_{N,0}\,,\la{com1}\\
&&[Q, \hat E^{\pm}_{N}]=\hat q(\mp\,\theta\,e^{i\theta N}\tau^{\pm})\,.\la{com2}
\ea
If we now define
\be\la{Vj}
\hat V=e^{iQ}\,,
\ee
we can show that the state $\hat V|0\rangle$ satisfies the properties \eqref{vj}. First, from \eqref{com1}
\be
\hat H^{3}_{0}(\hat V|0\rangle)=\hat V  (\hat V^{-1}\hat H^{3}_{0}\hat V) |0\rangle=
\hat V[-iQ,\hat H^{3}_{0}]|0\rangle=\frac 12(\hat V|0\rangle)\,.
\ee 
Furthermore, from \eqref{com1} and \eqref{com2} (by means of the Baker-Campbell-Hausdorff formula)
\ba
&&\hat V^{-1} \hat H^{3}_{\va N} \hat V= \hat H^{3}_{\va N}\,,\quad\text{for all}\;N\neq 0\,,\n\\
&&\hat V^{-1} \hat E^{\pm}_{\va N} \hat V= \hat q(e^{i(N\pm 1)\theta}\tau^\pm)=E^{\pm}_{\va N\pm 1}\,,
\ea 
which give
\ba\la{HEV}
&&\hat H^{3}_{\va N} (\hat V |0\rangle)=0\,~~N>0\n\\
&& \hat E^{+}_{\va N}(\hat V |0\rangle)=0\,~~N\geq 0\,.
\ea
In \eqref{HEV} we used the fact that  
\be
q^{i}_{\va N}|0\rangle=0~~~{\rm for}~~N\geq 0\,.
\ee
Therefore, we see that the vertex operation \eqref{Vj} creates a highest weight state associated with the spin-1/2 representation (for $k=1$ this is the only unitary irreducible representation) for the $\su(2)$ Kac-Moody algebra and is a primary field. As shown below, its conformal dimension is proportional to the square of the area carried by the puncture $p$. From \eqref{KM} we see that the zero modes $q^{i}_0$ constitute an $\su(2)$ Lie algebra and the full set $q^{i}_{\va N}$ furnishes an `affinisation' of $\su(2)$. 

As we saw above the representation theory of affine algebras shares many features with that of the Virasoro algebra. In particular, through the Sugarawa construction \eqref{L} the primary fields $V_j$ acting on the vacuum state $|0\rangle$ generate the highest weight states of the representation of both the algebras, namely Kac-Moody and Virasoro. This shows that the Kac-Moody primary fields are also Virasoro primary fields\footnote{Note however, that the inverse is in general not true: a Virasoro primary field can be a descendant in the Verma module associated to the Kac-Moody affine algebra.}. In the case of the affine algebra, these highest weight states \eqref{vacuum} provide a representation of the affine algebra
\be\la{qprim}
q^{i}_0|v_j\rangle=\tau^i_{\va (j)} |v_j\rangle,\qquad q^{i}_N|v_j\rangle=0~(N>0)\,,
\ee
where $\tau^i_{\va (j)}$ are the $\su(2)$ matrices in the spin-$j$ representation. From here, we can explicitly compute the action of the energy operator $\hat L_0$ on the highest weight state $|v_j\rangle$ for the $SU(2)$ Chern-Simons theory coupled to a source. From \eqref{L0}, the action of $\hat L_0$ on $|v_j\rangle$ is proportional to the quadratic Casimir operator of the $SU(2)$ irreducible $j$-representation
\be\la{Lv}
\hat L_0 |v_j\rangle = \frac{1}{k+2}\tau^i_{\va (j)} \tau^i_{\va (j)} |v_j\rangle = \frac{j(j+1)}{k+2}|v_j\rangle\,.
\ee
This action \eqref{Lv} is reminiscent of the flux operator acting on the spin network states of the kinematical Hilbert space $\Hk$ of LQG. This observation can be made more precise if we follow \cite{ABCK, SU(2)} in coupling the bulk and the boundary operators by identifying the $\su(2)$ generator $\hat S^i$ at the source $p$ with the flux operator associated with the 2-form $\Sigma^i=\epsilon^{i}_{\ jk} e^j\wedge e^k$ acting on $\Hk$, namely
\be
\epsilon^{ab}\hat{\Sigma}^i_{ab}(x) = 16 \pi \gamma G\hbar
  \delta(x,p) \hat{J}^i(p)\,,
\ee
where $\gamma$ is the Immirzi parameter of LQG. Then, one is led to define the vacuum representation of the Kac-Moody and Virasoro algebras in the quantum theory in terms of the $SU(2)$ spin network states associated to the links piercing the boundary at $p$s and coloured with spins $j_p$s. The action of $\hat J^i$ on a spin network link of colour $j$ is
\be
\hat{J}^i|j\rangle=\tau^i_{\va (j)} |j\rangle\,.
\ee
We have already seen above that the primary field $V_j$ creating the highest weight state \eqref{vacuum} has an interpretation in terms of the Wilson line along a link on the horizon terminating at the puncture $p$. Such an interpretation can be applied to the Wakimoto representation \cite{Wakimoto} of the vertex operator which allows one to incorporate structures of the LQG kinematical framework into the construction of the Verma module associated to the Kac-Moody (and hence, Virasoro) algebra. The Wakimoto representation provides a generalisation of the vertex representation introduced above for $k>1$. The fact that this free field representation can be regarded as an affine extension of the monomial representation of the $\su(2)$ Lie algebra  provides a setting which allows us to introduce spin network states. In order to see this, we introduce the Wakimoto free field representation in the next section.

\subsubsection{Free Field Representation}\la{sec:free-field}

The details of the free bosonic field representation of the affine algebras are given in Appendix \ref{Appendix1}.
On a complex plane the Laurent expansion of a free massless bosonic field is given by 
\be\la{phi-exp2}
\phi(z,\bar z)=\phi_0-ia_0\ln{z\bar z} +i\sum_{n\neq 0} \frac{1}{n}\left( a_n z^{-n}+ \bar a_{n}\bar z^{-n}\right)\,.
\ee
The conformal vacuum $|\alpha\rangle$, $\alpha\in\bf R$, of the theory is defined by
\be\la{vac3}
a_n|\alpha\rangle=\bar a_n|\alpha\rangle=0,\quad\text{for}\;n>0\qquad{\rm and}\qquad a_0|\alpha\rangle=\bar a_0|\alpha\rangle=\alpha |\alpha\rangle\,.
\ee
Such a one parameter family of states $|\alpha\rangle$ can be constructed from the absolute vacuum $|0\rangle$ by the vertex operator $V_\alpha(z,\bar z)=e^{i\alpha\phi(z,\bar z)}$, namely
\be
|\alpha\rangle=V_\alpha(0)|0\rangle\,.
\ee
Making use of the commutation relation $[\phi_0,a_n]=i\delta_{n0}$, it is easy to verify that the state $V_\alpha(0)|0\rangle$ satisfies \eqref{vac3}. From the operator product expansions (OPE) of these vertex operators with the energy-momentum tensor $T(z)$ \eqref{SET} it can be shown that the $V_\alpha(z,\bar z)$ are primary fields with holomorphic and anti-holomorphic conformal dimensions $h(\alpha)=\bar h(\alpha)=\alpha^2/2$. Furthermore, from the OPE of a string of vertex operators $V_{\alpha_i}$, $i=1,...,n$, with respective charges $\alpha_i$ it can be seen that the correlation function 
\be\la{correlator}
\langle V_{\alpha_1}(z_1, \bar z_1)V_{\alpha_2}(z_2, \bar z_2)\cdots V_{\alpha_n}(z_n, \bar z_n)\rangle=\prod_{i<j} (z_i-z_j)^{2\alpha_i\alpha_j}
\ee
vanishes unless the sum of charges vanishes, namely 
\be\la{neutrality}
\sum_i \alpha_i=0\,.
\ee
A simple way to see how this condition of the vanishing of charge arises is to impose the invariance of the correlation function \eqref{correlator} under the shift $\phi\rightarrow\phi+a$ where $a$ is a constant. This shift is a symmetry of the free boson Lagrangian in absence of a mass term. Under the shift, the correlation function picks up a phase $\exp{ia (\alpha_1+\cdots+\alpha_n)}$, from which \eqref{neutrality} follows by demanding the phase to be unity. 

As we have seen, the vertex operator $V_j$ which is used to construct the highest weight state of the Kac-Moody algebra at each puncture can also be interpreted as a holonomy (see \eqref{Vj}). In the context of $U(1)$ isolated horizons the  condition that the sum of the third component of the spin-$j$ colour associated with all the punctures must vanish follows from the boundary condition 
\be
F(A)=-\frac{2\pi}{k}\Sigma\,.
\ee
This relation between the boundary flux and the bulk geometry is at the origin of the intertwiner structure used to construct the horizon Hilbert space in the quantum theory and encode the correlations among all the punctures. The condition \eqref{neutrality} plays a similar role here in the context of free bosonic field representation. This shows how a free field interpretation of quantum isolated horizons boundary operators in LQG emerge. 

A free field representation of the $SU(2)$ Kac-Moody algebra at level $1$ can be obtained from the primary fields of conformal dimension 1 in the free bosonic theory. More precisely, defining
\be
H=i\partial \phi~~~~~~E^\pm=e^{\pm i \phi}\,,
\ee
it can be shown that their OPE correspond to the one of the Kac-Moody generators in the Cartan basis for $k=1$. Moreover, the Sugarawa construction (from \eqref{c} we find that $c=1$, in agreement with a single free bosonic theory) of the energy-momentum tensor reduces to that of a free boson \cite{CFT}. The theory then contains only one highest weight state given by the vertex operator $V_{1/2}=e^{i\phi/2}$.

For the case $k>1$, while trying to build the free field representation one finds that the Cartan generators $E^{\pm\alpha}$ now contain a term $e^{\pm i\alpha\phi/\sqrt{k}}$ in order to reproduce the correct OPE. But this is  problematic since such a vertex operator is no longer a weight 1 current, having conformal dimension $\alpha^2/2k^2$. This difficulty can be overcome by using the Wakimoto free filed representation.
Such representation provides an affine extension of the monomial basis of $SU(2)$ representing the states $|j,m\rangle$. More precisely, given the $\su(2)$ generators in the Chevalley basis $\{h_0, e_0, f_0\}$, they can be written in terms of creation and annihilation operators $\gamma_0,\beta_0$ respectively, with $[\beta_0,\gamma_0]=1$, and a momentum generator $\hat p$ canonically conjugate to a variable $\hat q$ ($[\hat p,\hat q]=-i$). A Fock space equivalent to the representation space of the spin-$j$ module of $\su(2)$ can then be constructed by repeated application of the creation operator $\gamma_0$ on a vacuum state characterized by its momentum eigenvalue $j$. Explicitly, one gets
\be\la{hef0}
h_0=2\hat p-2\gamma_0\beta_0~~~~~~~e_0=\beta_0~~~~~~~f_0=2\hat p\gamma_0-\gamma_0^2\beta_0\,.
\ee
An affine extension of this representation can then be constructed by regarding the $\su(2)$ generators as the zero modes of the affine generators. In other words, we can interpret the set $\{\gamma_0, \beta_0, \hat p, \hat q\}$ as the zero modes of appropriate free fields and replace them by the corresponding fields. In particular, $\hat q$ and $\hat p$ are identified with the zero modes of the free-bosonic field $\phi_0$ and $\pi_0$ respectively (see \eqref{phi-exp}). By replacing $\hat p$ by $i\sqrt{k+2}\,\partial\phi$, the affine extensions of the generators \eqref{hef0} are given by 
\begin{align}\la{hef}
&h(z)=i2\sqrt{k+2}\,\partial\phi(z)-2(\gamma\beta)(z),\quad e(z)=\beta(z),\n\\ 
&f(z)=i2\sqrt{k+2}(\partial\phi\gamma)(z)-(\beta\gamma^2)(z)-k\partial \gamma (z)\,,
\end{align}
where
\be
\beta(z)=\sum_n \beta_n z^{-n-1}~~~~~~~~\gamma(z)=\sum_n \gamma_n z^{-n}\,.
\ee
are two bosonic ghost fields of weight 1 and 0 respectively. 
It can be shown \cite{CFT} that the currents \eqref{hef} have the correct $SU(2)$ Kac-Moody OPE at level $k$. Moreover, the Suagrawa energy-momentum tensor in terms of these currents is equivalent to the one of a free bosonic field with a non-zero background charge $-1/2\sqrt{k+2}$ (modulo term coming from the ghost fields). Summing the contributions to the central charge coming from the bosonic field, the background charge and the ghost fields, one recovers the relation \eqref{c}.

The highest weight state for the Wakimoto representation of the $SU(2)$ Kac-Moody alegbra is given by 
\be\la{vjj}
v_{j,j}\equiv |j,j\rangle e^{\frac{i j \phi}{\sqrt{k+2}}}
\ee
and all the other states $v_{j,m}$ in the module are generated by successive OPE with the field $f(z)$, where
\be\la{vjm}
v_{j,m}\equiv |j,m\rangle e^{\frac{i j \phi}{\sqrt{k+2}}}\,,
\ee
where $|j,m\rangle$ represents the (ghost) monomial $\gamma^{j-m}$. Because of the zero weight of the $\gamma$ field, the $|j,m\rangle$ term does not contribute to the conformal dimension of the primaries \eqref{vjm}; it is the vertex operator term in \eqref{vjm} which provides a nonzero conformal dimension for the field in agreement with \eqref{Lv} once the background charge contribution is taken into account \cite{CFT}.

The action of the currents on $v_{j,m}$ is
\ba
&&f(z)v_{j,m}(w)\sim\frac{(j+m)v_{j,m-1}(w)}{z-w}\n\\
&&h(z)v_{j,m}(w)\sim\frac{(2m)v_{j,m}(w)}{z-w}\n\\
&&e(z)v_{j,m}(w)\sim\frac{(j-m)v_{j,m+1}(w)}{z-w}\,.
\ea 
If we now compare the two representations \eqref{Vj} and \eqref{vjj}, we see how one can associate a vertex operator at each puncture in terms of a bulk edge (ending at the puncture) of the spin network state and its tensor product with the holonomy of the boundary connection along an edge $e$, as depicted in Figure \ref{Disc2}. The zero mode of this boundary connection represents a geometric degrees of freedom (i.e. the flux of the gravitational electric field through the surface $H$ around the puncture) via the boundary condition \eqref{con2} and corresponds to the momentum of a free boson field. The higher modes in the connection are not geometrical, since they do not modify the flux, but can be associated with the bosonic modes $a_n$ in the expansion \eqref{phi-exp}. In other words, the affinization of the `gravitational' $SU(2)$ finite Lie algebra associated to the conformal symmetry at each puncture on the horizon introduces an infinite tower of new degrees of freedom that can be interpreted in terms 
of quanta of a free boson field (a representation in terms of fermonic fields is also possible \cite{CFT}).

\section{Partition Function}\la{sec:Entropy}

We now compute the partition function associated with the conformal family of states localised at the internal boundary $H$. Using \cite{Elitzur}, we map the CS-states on $D$ (in the limit $\epsilon_p\to 0$ where $\epsilon_p$ is the radius of the hole $H$ around the puncture $p$, the annulus $D$ becomes a disk with a source $S^i$) to the CS-theory on a torus having a modular parameter $\tau$ \footnote{A torus on the cylinder complex plane  may be defined by specifying two linearly independent lattice vectors on it and identifying points that differ by an integer combination of these vectors. These lattice vectors can be represented by 2 complex numbers $w_1$ and $w_2$, the periods of the lattice. The properties of a CFT on a torus depend only on the  modular parameter $\tau=w_2/w_1$. By identifying one of the periods with the one of the Euclidean time variable and the other one with the circumference of the hole $H$, we get $\tau=i/\epsilon_p$.} such that $\text{Im}\,\tau=1/\epsilon_p$. In the following, we 
consider the partition function of WZW model on a torus
\cite{CFT}
\be\la{Zp}
Z_{\va p}(\tau)=\tr\,e^{2\pi i\tau(L_0-\frac{c}{24})}e^{-2\pi i\bar\tau(\bar L_0-\frac{c}{24})}\,,
\ee 
where in all subsequent discussion we will drop the anti-holomorphic part to simplify the expressions. For a collection of $N$ non-interfering punctures the total partition function is proportional to the product
\be\la{Z}
Z(\tau)\propto\prod_{\va p}Z_{\va p}(\tau).\ee
%

The expression above for the partition function contains the geometrical dof corresponding to the microstates usually counted in the LQG literature. These appear in the sum over the primary fields of the theory, which have a representation in terms of holonomies, as shown above, and corresponds to the usual counting of conformal blocks in the previous literature. However, the expression \eqref{Z} now also includes new dof encoded in the sum over the secondary fields (the descendants) in each conformal block created by acting with combinations of the $q^i_{-N}$ generators on the primary fields. These extra charges have an interpretation in terms of non-geometric (matter) degrees of freedom, as we saw in Section \ref{sec:free-field}, and represent the novel feature of our analysis. This is the central observation of this paper.

The root of these additional structures lies in the modular invariance of $Z$ and the underlying affine Kac-Moody algebra. As pointed out in \cite{CFT}, the partition function \eqref{Z} expressed as a sum of the $SU(2)$ Kac-Moody characters associated with the irreducible representations labelled by spin $j$, has a dependence on an angular coordinate $\theta$ labelling the maximal torus of $SU(2)$. This is in addition to its dependence on the modular parameter $\tau$. In other words, the true characters are $\chi(\tau,\theta)$ and the partition function is   
\be
Z(\tau,\theta)=\prod_{\va p}\sum_{j_p=0}^{k/2}\chi^k_{j_p}(\tau, \theta)\,.\ee
Using the Kac-Weyl formula, the $SU(2)$ character for the spin-$j$ representation at level $k$ and $\theta=\pi$ is given by (see Appendix \ref{Appendix2} for details; also \cite{GKO} and \cite{Wassermann})
\be\la{chi}
\chi^k_{j}(\tau,\pi)= \tr_{j,k}{[q^{L_{\va 0}-\frac{c}{24}}e^{ i\pi \hat H_{\va 0}^3}]}=\frac{q^{\frac{(2j+1)^2}{4(k+2)}}\sum_{n\in\Z}(-1)^{2j}(2j+1+(k+2)n)q^{(k+2)n^2+(2j+1)n}}{\prod_{n=1}^\infty(1-q^n)^3}\,,
\ee
where $q=e^{-2\pi\text{Im}\tau}$. The choice $\theta=\pi$ in evaluating the partition function is the correct one in order to take into account all the new degrees of freedom \cite{CFT}, as explained in Appendix \ref{Appendix2}. In the double limit $\epsilon_p\to 0$ (or $q\to 0$) and $k\to\infty$ such that $\epsilon_pk$ approaches a finite value, the denominator in \eqref{chi} being independent of $j$ approaches $1$. On the other hand, the sum over $n$ in the numerator is such that all terms with $n\neq 0$ are exponentially suppressed by the $n^2$ term in the exponent and only the $n=0$ term survives in the limit. Finally, we arrive at the partition function 
\begin{align}\la{Zcan}
Z=\prod_{p=1}^N \sum_{j_p=0}^{k/2} (-1)^{2j_p}(2j_p+1)\,q^{\frac{j_p(j_p+1)}{k}}\,.
\end{align}
We can thus rewrite the partition function in the following form
\be\la{Zcan2}
Z=\prod_{p=1}^N \sum_{j_p=0}^{k/2}q^{\Delta_{j_p}}\rho( j_p)\,,
\ee
where $\Delta_j=j(j+1)/(k+2)$ is the eigenvalue of $L_0$ for the highest weight state \eqref{Lv} and 
\be\la{rho}
\rho(j_p)=(2j_p+1)e^{2\pi i j_p}\sim(2j_p+1)\exp{(a_p/4\ell^2_{\va P})}\,
\ee
is an additional degeneracy factor. In fact, in a general CFT $\tr\,q^L_0=\sum_j\rho(j)q^{\Delta_j}$, where $\rho(j)$ characterizes the number of degenerate states that occur at a fixed energy level $\Delta_j$. In passing to the r.h.s of \eqref{rho} we have used the LQG spectrum of the area operator at the given puncture $p$, namely\footnote{In the literature one usually finds an expression for the area spectrum in terms of the square root of the quadratic $SU(2)$ Casimir operator evaluated in the representation $j$ of the form $C(j)=j(j+1)$; however, there exists also a different regularization in which the area operator is linear in $j$, i.e. $C(j)=j^2$. The advantages of using this latter form of area spectrum in the context of BH entropy calculation have been studied in \cite{Barbero}. The two spectra obviously coincide in the large spin limit. The expression \eqref{ap} can be seen either as this alternative regularization or as the large spin limit of the usual one: the choice does not affect our 
discussion.}
\be\la{ap}
a_p = 8 \pi \ell^2_{\va P} \gamma j_p 
\ee
and taken $\gamma=i$; this is how an imaginary Barbero-Immirzi parameter enters our derivation.

Thus, \eqref{Zcan} shows that the CFT partition function contains a degeneracy factor at each puncture. In the context of LQG this additional factor seems to be missing in the previous calculations (see also the recent analyses of  \cite{Han, Geiller-Noui} for the intorduction of a similar degeneracy factor from the formula of the Chern-Simons Hilbert space dimension). 
Therefore, for the choice $\gamma=i$, corresponding to the self-dual connection, the degeneracy factor \eqref{rho} reproduces Bekenstein's holographic bound.

Therefore, by taking into account the new degrees of freedom contained in the Kac-Moody algebra, one can apply quantum statistics to punctures which led to a result for the entropy that reproduces the semiclassical formula \cite{GNP} without the large quantum gravity corrections associated with the number of punctures as quantum hair \cite{AP, Radiation, Radiation2}. Moreover, our analysis is consistent with and provides further evidence to the recent observations of the special role played by an imaginary Barbero-Immirzi parameter in recovering the right numerical factor in front of the entropy-area relation \cite{Complex, Temp}, as reported in the Introduction. In light of the possible interpretation of the extra charges $q_{\va N}$ \eqref{qH2} in terms of free matter d.o.f. pointed out above and the result of \cite{Temp} on the equivalence between statistical and entanglement entropy, this new contribution to the IH state degeneracy could then be interpreted as an entanglement entropy contribution coming from matter fields near the horizon, as originally assumed in \cite{GNP}.


Another interesting aspect of \eqref{Zcan} is that it may be possible to interpret $Z$ as a partition function of a gas of punctures. Of course, care is needed because the Barbero-Immirzi parameter involved here is imaginary, $\gamma=i$. However, if we also complexify the CS-level $k$ (see e.g. \cite{Han}), by choosing $\epsilon_pk=\sqrt{j_p(j_p+1)}/i$ we then formally get (compare with \cite{GP})
\be
Z=\prod_{p=1}^N \sum_{j_p=0}^{k/2}e^{-\beta E_{j_p}}\rho(j_p)\,,
\ee
where $\beta$ is the local inverse Unruh temperature and $E_{j_p}$ is the appropriate local energy associated with a single puncture obtained from the area spectrum at the puncture. However, because $E_j$ involves $\gamma=i$, the exponent is oscillatory and $Z$ looses its standard interpretation as a partition function. Then one could follow the analytic continuation of \cite{Han} and eventually interpret $Z$ as a genuine partition function.

\section{Conclusions and Comments}\la{sec:Conclusions}

By properly taking into account the regularization in the form of introducing a new boundary for each puncture that is necessary in order to have a well defined quantization of Chern-Simons theory on a punctured 2-sphere, we have defined the Hilbert space of a quantum IH. An infinite set of charges living on the new boundary around each point can then be associated with each puncture where a bulk link pierces the horizon. These can be shown to satisfy a Kac-Moody algebra, 
to which a Virasoro algebra can be canonically associated, representing the algebra of diffeomorphisms on the circle and encoding a local conformal symmetry. Due to the presence of a central charge in the Virasoro algebra, only half of the new charges can still be regarded as gauge d.o.f. while the rest turns into physical d.o.f. as a consequence of the partial breaking of gauge (diffeo) symmetry due to the presence of new boundaries on the horizon. 

By imposing the IH boundary conditions for the gravitational fluxes defined on a disk with punctures, these are shown to correspond to the zero modes of the Kac-Moody algebra defined on the boundaries of the disk. In this way, the gravitational degrees of freedom can be encoded in some conformal charges, suggesting a CFT/gravity correspondence as long as the $q_0$ observables are involved. 
 Moreover, the LQG spin network states represent the highest weight states which provide unitary irreducible representations of the Kac-Moody and Virasoro algebras. In this sense, the standard picture of quantum IH is contained in our conformal description. At the same time though, the role of the higher modes on the CFT side in such a correspondence with the LQG degrees of freedom is less clear. An intriguing scenario comes from the observation that  fields become effectively massless at the horizon of a black hole and have conformal symmetry \cite{Birmingham}; in Section \ref{sec:free-field} we showed how the extra charges associated with the higher modes have a representation in terms of free matter fields. Therefore, the unified CFT description might provide an alternative way to couple matter  degrees of freedom in the LQG framework, at least on the surface of an isolated horizon. In this perspective then, the correspondence would involve CFT degrees of freedom on one side and gravity+matter degrees of 
freedom in the LQG description on the other. This speculative idea requires further investigation, of course.
 
What is clear is that the higher Kac-Moody charges represent new physical degrees of freedom associated to the presence of new boundaries on the IH and are ultimately responsible for the disappearance of the entropy contribution associated to the number of punctures. In fact, by mapping the CS Hilbert space on a disk with punctures to that of a WZW theory on a torus, we can use standard results of 
the CFT partition function in terms of the Kac-Weyl characters to show that a new degeneracy factor satisfying the Bekenstein's bound appears. Such degeneracy factor in the partition function \eqref{Zcan2} also provides further evidence to the recent discovery of the special role played by an imaginary Barbero-Immirzi parameter in recovering the right numerical factor in front of the entropy-area law \cite{Complex, Temp}.

Let us conclude with some comments. The dynamics induced by the energy operator $\hat L_0$ has a nice interpretation as self-interaction of holonomies. More precisely, the regularisation procedure represented by the boundary circle $\partial H$ around each puncture is the analog of framing of links in $\R^3$; while computing the expectation value of a Wilson line $\langle W(C)\rangle$ in three dimensions using the Chern-Simons path integral, a shift of the framing of the link $C$ by $\tau$ units induces a phase given by the conformal dimension \cite{Witten, Guadagnini}, namely $\langle W_j(C)\rangle\rightarrow e^{2\pi i\tau\Delta_j}\langle W_j(C)\rangle$. From the physical viewpoint, the notion of particle dynamics in terms of change of framing of the thickened Wilson-line representing the particle trajectory is consistent with the IH boundary conditions: since the system is isolated, punctures cannot interact with each other and the only allowed interactions are  self-interactions.
 
For the CFT partition function \eqref{Zp} on a torus with a modular parameter $\tau=i/\epsilon_p$ the modular invariance ($\tau\rightarrow-1/\tau$) and the periodic boundary conditions imply a notion of finite inverse temperature $\beta=\epsilon_p$ \cite{Cardy2}. Interestingly, the fact that the relevant periodicity for the definition of CFT temperature is the one of the boundary circle picked out by the Virasoro algebra (instead of imaginary time) matches  the notion of IH geometric temperature associated to the the periodicity of the rotational symmetry around the puncture derived in \cite{Temp}. The correspondence between these two notions of horizon temperature could be traced back to the  isomorphism between the Lorentz algebra symmetry in four dimensions (on which the derivation of \cite{Temp} was based) and the conformal algebra in two dimensions.
 The implications of repeating the analysis of  \cite{Temp} by replacing  the internal boost generator with the Virasoro energy operator \eqref{Lv} as time evolution Hamiltonian is left for future investigation. 

Given the local nature of our treatment, the result \eqref{Zcan} can be extended to non-spherically symmetric IH as long as the boundary conditions coupling the bulk and the horizon theories can be recast in the form $F(A)\propto\Sigma$ for some $SU(2)$ connection $A$ \cite{genericIH, Engle}.

We end with a brief discussion of the microcanonical entropy of a CFT computed from the Cardy formula \cite{Cardy}. Cardy's result for the asymptotic density of states of  a CFT depends only on the central charge and the eigenvalues of the Virasoro generator $\hat L_0$. In our case, the first quantity is given by \eqref{c}, which for $\g=\su(2)$ gives $c=3k/(k+2)$; in the large $k$ limit $c=3$. The second quantity has been derived in \eqref{Lv}. However, in order for the Cardy's asymptotic formula to be valid, one must have $L_0\geq c$. With our assumption of a large value for the CS level corresponding to a macroscopic black hole, Cardy's formula can thus be applied only in the limit of large spins associated to the punctures, i.e. $j\sim k$. Even in this approximation, after summing over all the punctures it is not clear that a linear growth of entropy with the IH area is recovered, let alone the right numerical factor. Therefore, application of Cardy's formula in our context does not seem to be as 
straightforward and a detailed analysis is necessary.

\section*{Acknowledgement}

A.G. would like to thank the kind hospitality of the Albert Einstein Institute (Max Planck Institute for Gravitational Physics), Potsdam, Germany, during the summer of 2013 where the work was initiated. A.G. would also like to acknowledge the Department of Atomic Energy, Govt. of India, for funding and support.


\appendix

\begin{appendix}

\section{Free Boson}\la{Appendix1}

Here we closely follow \cite{CFT}. Consider a massless free bosonic field $\phi(x,t)$ on a cylinder of circumference $L$, satisfying the boundary condition
\be\la{phi-L}
\phi(x,t)=\phi(x+L,t)\,.
\ee
Its Hamiltonian reads
\be\la{Ham}
H=\frac{2\pi}{L}\sum_n\left(\pi_n\pi_{-n}+\left(\frac{n}{2}\right)^2\phi_n\phi_{-n}\right)\,,
\ee
where $\pi_n=\frac L{4\pi}\dot\phi_{-n}$ is the conjugate momentum of $\phi_n$ and
\be
[\phi_n,\pi_m]=i\delta_{nm}\,.
\ee
Notice that $\phi^\dagger_n=\phi_{-n}$ and $\pi^\dagger_n=\pi_{-n}$. The Hamiltonian \eqref{Ham} represents an infinite sum of decoupled harmonic oscillators of frequencies $\omega_n=|n|/2$. The masslessness is equivalent to the vanishing of the zero mode $\omega_0=0$. This, together with the boundary condition \eqref{phi-L}, implies conformal invariance of the theory with a central charge $c=1$. 

If we now introduce the following creation and annihilation operators for the $n\neq 0$ modes
\be
a_n=-i\frac{n}{2}\phi_n+\pi_{-n},\qquad\bar a_n=-i\frac{n}{2}\phi_{-n}+\pi_n,\quad\text{for}\;n\neq0
\ee
they satisfy,
\be
[a_n,a_m]=n\delta_{n+m},\quad[a_n,\bar a_m]=0,\quad[\bar a_n,\bar a_m]=n\delta_{n+m},
\ee
and the Hamiltonian becomes
\be\la{H-bos}
H=\frac{2\pi}{L} \left(\pi_0^2+\frac 12\sum_{n\neq0} a_{-n}a_n+\bar a_{-n}\bar a_n\right)\,.
\ee
From the commutation relations above we have that
\be
[H,a_{-n}]=\frac{2\pi}{L} n a_{-n}\,,
\ee
which implies $a_{-n} (n>0)$ acting on an eigenstate of $H$ with energy eigenvalue $E$ produces another eigenstate with energy eigenvalue $E+2n\pi/L$.

The Fourier expansion of the field $\phi(x,t)$ at arbitrary time then becomes 
\be
\phi(x,t)=\phi_0+\frac{4\pi}{L}\pi_0 t+i\sum_{n\neq 0} \frac{1}{n}\left(a_ne^{2\pi i n(x-t)/L}- \bar a_{-n}e^{2\pi i n(x+t)/L}\right)\,;
\ee
by going to the Euclidean time $\tau=i t$ and introducing the planar conformal coordinates
\be
z=e^{2\pi (\tau-ix)/L},\qquad\bar z=e^{2\pi (\tau+ix)/L}\,,
\ee
we obtain the expansion
\be\la{phi-exp}
\phi(z,\bar z)=\phi_0-i\pi_0\ln{z\bar z} +i\sum_{n\neq 0} \frac{1}{n}\left( a_n z^{-n}+ \bar a_{n}\bar z^{-n}\right)\,.
\ee
If we consider only the holomorphic part and introduce the operator $a_0=\bar a_0=\pi_0$, we can express the derivative $\partial \phi=\partial_z\phi$ as
\be
i\partial \phi=\sum_{n\in\Z} a_n z^{-n-1}\,.
\ee
Comparing with \eqref{primary}, we see that the derivative field $\partial\phi$ is a primary field with conformal dimension $h=1$. 

The holomorphic energy-momentum tensor is given by
\be\la{SET}
T(z)=-\frac{1}{2}:\partial\phi\partial\phi:\,= \frac{1}{2}\sum_{n,m\in \Z}z^{-n-m-2}:a_n a_m:\,,
\ee
where the normal ordering ensures that the vacuum expectation value of of the energy-momentum tensor vanishes. The conformal vacuum $|\alpha\rangle$ is then defined by
\be\la{vac1}
a_n|\alpha\rangle=\bar a_n|\alpha\rangle=0,\quad\text{for all}\;n>0,
\ee
and it is an eigenstate of the zero mode $a_0=\bar a_0$
\be\la{vac2}
a_0|\alpha\rangle=\bar a_0|\alpha\rangle=\alpha |\alpha\rangle\,.
\ee
Since $a_0$ commutes with all the $a_n,\bar a_n$, these operators cannot change the eigenvalue of $a_0$.

In terms of the modes expansion \eqref{T-exp}, the expression \eqref{SET} implies
\ba\la{L-bos}
&&L_n=\frac{1}{2}\sum_{m\in\Z}:a_{n-m}a_m:\,,\qquad n\neq 0\n\\
&&L_0=\frac 12\sum_{n\neq 0}:a_{-n}a_n:+\frac{1}{2}a_0^2,
\ea
and similarly, for the anti-holomorphic modes. The Hamiltonian \eqref{H-bos} then takes the form
\be
H=\frac{2\pi}{L} \left(L_0+\bar L_0\right)\,.
\ee
From the zero mode \eqref{L-bos}, the vacuum $|\alpha\rangle$ has conformal dimension $\alpha^2/2$. The elements of the Fock space are obtained by acting on $|\alpha\rangle$ with $a_{-n}, \bar a_{-n}$ with $n>0$.

\section{Kac Character Formula}\la{Appendix2}

Here we outline the derivation of \eqref{chi}. To do so we make use of the formula for the normalised character of the irreducible positive energy spin-$j$ representation of affine $SU(2)$ of level $k$ given in \cite{Wassermann}, 
\be
\tr_{j,k}{[q^{\hat L_{\va 0}-\frac{c}{24}}z^{\hat H_{\va 0}^3}]}= \frac{\Theta_{2j+1,k+2}(q,z)-\Theta_{-2j-1,k+2}(q,z)}{\Delta(q,z)}\,,
\ee
where  the function $\Theta$ and the denominator $\Delta$ are given by
\begin{align}
&\Theta_{\ell,m}(q,z)=\sum_nz^{2m n}q^{mn^2}\,,\quad\text{where}\;n\in\frac{\ell}{2m}+\Z\,\\
&\Delta(q,z)=\Theta_{1,2}(q,z)-\Theta_{-1,2}(q,z)=(z-z^{-1})\prod_{n>0}(1-q^n)(1-z^2q^n)(1-z^{-2}q^n)\,.
\end{align}
Hence, we have
\begin{align}
\tr_{j,k}{[q^{\hat L_{\va 0}-\frac{c}{24}}z^{\hat H_{\va 0}^3}]}&= 
\frac{\left(\sum_{n\in\frac{2j+1}{2(k+2)}+\Z}-\sum_{n\in\frac{-(2j+1)}{2(k+2)}+\Z}\right)z^{2(k+2)n}q^{(k+2)n^2}}{(z-z^{-1})\prod_{n>0}(1-q^n)(1-z^2q^n)(1-z^{-2}q^n)}\nonumber\\
&=\frac{\sum_{n\in\Z}q^{\frac{(2j+1)^2}{4(k+2)}}\left(z^{2j+1}z^{2(k+2)n}q^{n(2j+1)}-z^{-(2j+1)}z^{2(k+2)n}q^{-n(2j+1)}\right)q^{n^2(k+2)}}{(z-z^{-1})\prod_{n>0}(1-q^n)(1-z^2q^n)(1-z^{-2}q^n)}\nonumber\\
&=\frac{q^{\frac{(2j+1)^2}{4(k+2)}}\sum_{n\in\Z}\left( \frac{z^{2j+1+2(k+2)n}-z^{-(2j+1+2(k+2)n)}}{(z-z^{-1})}\right)q^{n(2j+1)+n^2(k+2)}}{\prod_{n>0}(1-q^n)(1-z^2q^n)(1-z^{-2}q^n)}\,.
\end{align}
Since an element of the maximal torus of $SU(2)$ is parametrised by $\left(\begin{smallmatrix} z & 0\\ 0 & 1/z \end{smallmatrix}\right)$, we can choose $z=e^{i\theta}$ and 
\bee
\tr_{j,k}{[q^{\hat L_{\va 0}-\frac{c}{24}}e^{i\theta\hat H_{\va 0}^3}]}=\frac{q^{\frac{(2j+1)^2}{4(k+2)}}\sum_{n\in\Z}\frac{\sin{\theta(2j+1+2(k+2)n)}}{\sin{\theta}}q^{n(2j+1)+n^2(k+2)}}{\prod_{n>0}(1-q^n)(1-e^{2i\theta}q^n)(1-e^{-2i\theta}q^n)}\,.
\eee
Recall that $\theta=0,\pi$ are the only singular elements of the torus since their normaliser is the entire $SU(2)$ group. The choice $\theta=0$ corresponds to the so-called {\it specialized characters}, in which the affine Lie algebra structure behind the Virasoro algebra is neglected. Therefore, in order to include all the new degrees of freedom, the correct choice is $\theta=\pi$ \cite{CFT}, which gives
\bee
\tr_{j,k}{[q^{\hat L_{\va 0}-\frac{c}{24}}e^{i\pi\hat H_{\va 0}^3}]}=\frac{q^{\frac{(2j+1)^2}{4(k+2)}}\sum_{n\in\Z}(-1)^{2j}(2j+1+2(k+2)n)q^{n(2j+1)+n^2(k+2)}}{\prod_{n>0}(1-q^n)^3}\,,
\eee
reproducing our \eqref{chi}.

\end{appendix}


\end{document}